\documentclass[prd,twocolumn,nofootinbib,notitlepage,10pt,aps]{revtex4-1}
\usepackage[utf8]{inputenc}
\usepackage{cancel}
\usepackage{xcolor}
\usepackage{latexsym}
\usepackage{amsmath}
\usepackage{amssymb}
\usepackage{bbm}

\usepackage{hyperref}
\hypersetup{
    colorlinks,
    linkcolor={violet},
    citecolor={teal},
    urlcolor={teal}
}
\usepackage{orcidlink}

\usepackage{ulem}
\usepackage{pdfsync}
\usepackage{epsfig}
\usepackage{epstopdf}
\usepackage{subfigure}
\usepackage{color}
\usepackage{comment}
\usepackage{slashed}
\usepackage{physics}

\def\beq{\begin{equation}}
\def\eeq{\end{equation}}
\def\baq{\begin{eqnarray}}
\def\eaq{\end{eqnarray}}

\newcommand{\be}{\begin{equation}} % only untightened
\newcommand{\ee}{\end{equation}}
\newcommand{\bea}{\begin{eqnarray}} % only untightened
\newcommand{\eea}{\end{eqnarray}}

\newcommand{\bmp}{\noindent\begin{minipage}{16cm}}
\newcommand{\emp}{\end{minipage}\vskip 7mm} % 7mm untightened
\def\lsim{\mathrel{\raise.3ex\hbox{$<$\kern-.75em\lower1ex\hbox{$\sim$}}}}
\def\gsim{\mathrel{\raise.3ex\hbox{$>$\kern-.75em\lower1ex\hbox{$\sim$}}}}

\newcommand{\intron}[1]{}%{#1}

\def\MP{M_{\rm P}}

\begin{document}

%\title{Higgs-inflation and ACT}
%\title{Radiatively-corrected Higgs inflation in light of ACT}
\title{Has ACT measured radiative corrections to the tree-level Higgs-like inflation?}

\author{Ioannis D. Gialamas\orcidlink{0000-0002-2957-5276}}
\email{ioannis.gialamas@kbfi.ee}
 \author{Alexandros Karam\orcidlink{0000-0002-0582-8996}}
\email{alexandros.karam@kbfi.ee}
 \author{Antonio Racioppi\orcidlink{0000-0003-4825-0941}}
 \email{antonio.racioppi@kbfi.ee}
 \author{Martti Raidal\orcidlink{0000-0001-7040-9491}}
 \email{martti.raidal@cern.ch}

\affiliation{National Institute of Chemical Physics and Biophysics, \\
R\"avala 10, 10143 Tallinn, Estonia}

\begin{abstract}

Starobinsky inflation and nonminimally coupled (NMC) Higgs inflation have been among the most favored models of the early Universe, as their predictions for the scalar spectral index $n_s$ and tensor-to-scalar ratio $r$ fall comfortably within the constraints set by Planck and BICEP/Keck. However, new results from the Atacama Cosmology Telescope (ACT) suggest a preference for higher values of $n_s$, introducing tension with the simplest realizations of these models. In this work, being agnostic about the nature of the inflaton, we show that incorporating one-loop corrections to a quartic NMC inflationary scenario leads to a shift in the predicted value of $n_s$, which brings NMC inflation into better agreement with ACT observations. 
%Remarkably, we find that this can be achieved with nonminimal couplings $\xi < 1$, in contrast to the large values typically required in conventional Higgs inflation, thereby pushing any unitarity-violation scale above the Planck scale. 
The effect is even more significant when the model is formulated in the Palatini approach, where the modified field-space structure naturally enhances deviations from the metric case. These findings highlight the importance of quantum corrections and gravitational degrees of freedom in refining inflationary predictions in light of new data.  

\end{abstract}

%%%%%%%%%%%%%%%%%%%%%%%%%%%%%%%%%%%%%%%%%%%%%%%%%%%%%%%%%%%%%%%%%%%%%%%%%%%%%%%%%%%%%%%%%%%%%%%%%%%%
% DOCUMENT
%
\maketitle
%%%%%%%%%%%%%%%%%%%%%%%%%%%%%%%%%%%%%%%%%%%%%%%%%%%%%%%%%%%%%%%%%%%%%%%%%%%%%%%%%%%%%%%%%%%%%%%%%%%%

%%%%%%%%%%%%%%%%%%%%%%%%%%%%%%%%%%%%%%%%%%%%%%%%%%%%%%%%%%%%%%%%%%%%%%%%%%%%%%%%%%%%%%%%%%%%%%%%%%%%
%%%%%%%%%%%%%%%%%%%%%%%%%%%%%%%%%%%%%%%%%%%%%%%%%%%%%%%%%%%%%%%%%%%%%%%%%%%%%%%%%%%%%%%%%%%%%%%%%%%%
%
\section{Introduction}
Standard nonminimally coupled (NMC) tree-level Higgs inflation~\cite{Bezrukov:2007ep} and Starobinsky inflation~\cite{Starobinsky:1980te} have long been considered among the most successful models of cosmic inflation~\cite{Kazanas:1980tx, Sato:1981qmu, Guth:1980zm, Linde:1981mu}. Their predictions for the spectral index, $n_s$, and the tensor-to-scalar ratio, $r$, have consistently fallen well within the observationally allowed regions. However, the latest data release from the Atacama Cosmology Telescope (ACT)~\cite{ACT:2025fju, ACT:2025tim}, when combined with cosmic microwave background (CMB) measurements from BICEP/Keck (BK)~\cite{BICEP:2021xfz} and Planck~\cite{Planck:2018jri}, along with the first-year DESI measurements of baryon acoustic oscillations (BAO)~\cite{DESI:2024mwx}, introduce significant shifts in these constraints. In particular, the combination of Planck, ACT, and DESI (P-ACT-LB) leaves $r$ largely unchanged but notably revises the predicted value of the spectral index to $n_s = 0.9743 \pm 0.0034$. This updated value challenges the viability of NMC and Starobinsky inflation, as only a small fraction of the standard 50-60 $e$-folds period remains within the $2\sigma$ allowed region. In this regard, the latest observations have led to a reassessment of several inflationary models to ensure compatibility with the new data~\cite{Kallosh:2025rni, Aoki:2025wld, Berera:2025vsu, Brahma:2025dio, Dioguardi:2025vci}.

In this paper, without assuming a specific underlying particle content or inflaton scalar, we incorporate the radiative corrections~(e.g., Refs~~\cite{Marzola:2016xgb, Racioppi:2018zoy, Racioppi:2019jsp} and references therein) that inevitably will appear in a {\it Higgs-like} inflation scenario. By {\it Higgs-like} inflaton, we refer to a scalar whose potential during inflation is dominated by a self-quartic term and possesses only scale-invariant interactions with the particles in the ultraviolet (UV) complete theory i.e. only dimensionless couplings. The only exception to the last feature comes from the gravitational sector, where only the nonminimal coupling to gravity  $\xi \phi^2 R$ respects the last criterion. These statements are to be understood in the Jordan frame, where quantum corrections are incorporated. 

We show that the aforementioned corrections lead to inflationary predictions that remain well within current observational bounds.  Moreover, we find that smaller values of the nonminimal coupling, $\xi$, can still be viable. Finally, we explore alternative formulations of gravity and demonstrate that the Palatini formulation  (e.g.,~\cite{Koivisto:2005yc, Bauer:2008zj, Gialamas:2023flv} and references therein) offers improved agreement with the latest observational data compared to the metric formulation. As a result, we suggest that the ACT may have detected signatures of radiative corrections to the inflationary potential -- a possibility that warrants further investigation and must be tested by future experiments.

%%%%%%%%%%%%%%%%%%%%%%%%%%%%%%%%%%%%%%%%%%%%%%%%%%%%%%%%%%%%%%%%%%%%%%%%%%%%%%%%%%%%%%%%%%%%%%%

\section{Model}
\label{sec:the_model}
Following our definition of Higgs-like, we consider a theory involving a nonminimally coupled scalar field, $\phi$, specified by the action
\begin{equation}
\mathcal{S} = \int\dd^4x \sqrt{-g}\left(\frac{M_P^2 + \xi\phi^2}{2} R(g,\Gamma) + \frac{(\partial_\mu \phi)^2}{2}  - V_{\rm eff}(\phi) \right) ,
\label{eq:action1}
\end{equation}
where $M_P\simeq 2.4\times10^{18}$ GeV is the reduced Planck mass, $ V_{\rm eff}(\phi)$ is the 1-loop-corrected scalar potential, and $\xi$ is its nonminimal coupling to gravity which we assume to be constant.  The Ricci scalar $R$ is constructed from a connection $\Gamma$, which, in our analysis, can either be the Levi-Civita connection (in metric gravity) or an independent connection (in Palatini gravity) (e.g., Refs~\cite{Koivisto:2005yc, Bauer:2008zj, Gialamas:2023flv} and references therein). 

We focus on a 1-loop effective quartic scalar potential, 
\begin{equation}
  V_\text{eff}(\phi,\mu) = \frac{\lambda_\text{eff}(\phi)}{4} \phi^4\,,
  \label{eq:Veff}
\end{equation}
where $\lambda_\text{eff}$ is parametrized as \cite{Marzola:2016xgb, Racioppi:2018zoy, Racioppi:2019jsp}
\begin{equation}
  \lambda_{\rm eff}(\phi) \simeq \lambda(M_P) \left[1 + \delta(M_P) \ln\left(\frac{\phi}{M_P}\right)\right]\,,
  \label{eq:lrun}
\end{equation} 
with $\delta$ being the relative loop correction. Such corrections\footnote{ We work under the reasonable assumption that the mass terms of the full particle spectrum (inflaton included) are much below the inflationary scale and therefore, numerically irrelevant in Eqs.~\eqref{eq:Veff} and~\eqref{eq:lrun} for the treatment of the corresponding slow-roll dynamics and radiative corrections. The exact value of the mass terms might have an influence for the details of the dynamics out of the slow-roll regime, like the reheating mechanism, which is anyhow above the scope of the present work (see also Sec.~\ref{sec:inflation}). 
Moreover, it has been proven that the running of $\xi$ is subdominant (e.g., Refs~\cite{Marzola:2016xgb, Racioppi:2018zoy, Racioppi:2019jsp} and references therein), and is therefore ignored as well in our analysis. More details on the treatment of radiative corrections are given also in Appendix~\ref{Appendix:CW}.} to $\lambda$ originate from the full particle spectrum of a UV-complete theory. However, we remain agnostic about the exact particle content of the theory (apart from the assumptions of only classically scale-invariant interactions in the Jordan frame) in order to provide a model-independent study\footnote{Radiative corrections in the context of standard model (SM) Higgs inflation have been extensively studied in both the metric~(e.g. Refs.~\cite{DeSimone:2008ei, Barvinsky:2008ia, Barvinsky:2009ii, Barbon:2009ya, Bezrukov:2009db, Bezrukov:2010jz, Bezrukov:2012sa, Bezrukov:2014bra, Bezrukov:2014ipa, George:2015nza, Fumagalli:2016lls, Bezrukov:2017dyv, Markkanen:2018bfx, Okada:2015lia} and references therein) and Palatini formulations~(e.g. Refs.~\cite{Rasanen:2017ivk, Shaposhnikov:2020fdv, Enckell:2020lvn, Poisson:2023tja} and references therein). In our analysis, we consider general radiative corrections to nonminimally coupled quartic models, which merely resemble the conventional Higgs inflation scenario, i.e.,~only dimensionless couplings without necessarily identifying the inflaton with the SM Higgs field.}.
The analysis of inflationary observables is simplified in the Einstein frame, obtained through a Weyl rescaling of the metric tensor of the form $\Bar{g}_{\mu\nu} = (M_P^2+\xi\phi^2)/M_P^2 \, g_{\mu\nu}$. Applying this rescaling to the action~\eqref{eq:action1} and performing the following field redefinition:
\begin{equation}
\frac{\dd \chi}{\dd \phi} = \sqrt{ \frac{6\xi ^2 \phi ^2 M_P^2}{\left(M_P^2+\xi  \phi
   ^2\right)^2} \varepsilon + \frac{M_P^2}{M_P^2+\xi  \phi^2} } \, ,  
  \label{eq:dphiE}
\end{equation}
where $\varepsilon = 1$ in the metric and $\varepsilon = 0$ in the Palatini formulation, we obtain
\begin{equation}
\mathcal{S} = \int \dd^4x \sqrt{- \Bar{g}} \Bigg[ \frac{M_P^2}{2} R  + 
  \frac{(\partial_\mu \chi)^{2}}{2} - U(\chi) \Bigg] \, ,
   \label{eq:EF_lag}
\end{equation}
where the Einstein frame potential is given by
\begin{equation}
U(\chi) = \frac{M_P^4 \, V_\text{eff}(\phi(\chi))}{\left(M_P^2+\xi  \phi^2(\chi)\right)^2}\,.
\label{eq:U}
\end{equation}
Therefore, since the scalar field in~\eqref{eq:EF_lag} is canonically normalized, the only difference between the two formulations is given by the functional form of $\phi(\chi)$ in the Einstein frame potential~\eqref{eq:U}.

\section{Inflation} 
\label{sec:inflation}

%%%%%%%%%%%%%%%%%%%%%%%%%%%%%%%%%%%%%%%%%%%%%%%%%%%%%%%%
%%%%%%%%%%%%%%%%%%%%%%%%%%%%%%%%%%%%%%%%%%%%%%%%%%%%%%%%
\begin{figure*}[t!]
    \centering
    \includegraphics[width=\textwidth]{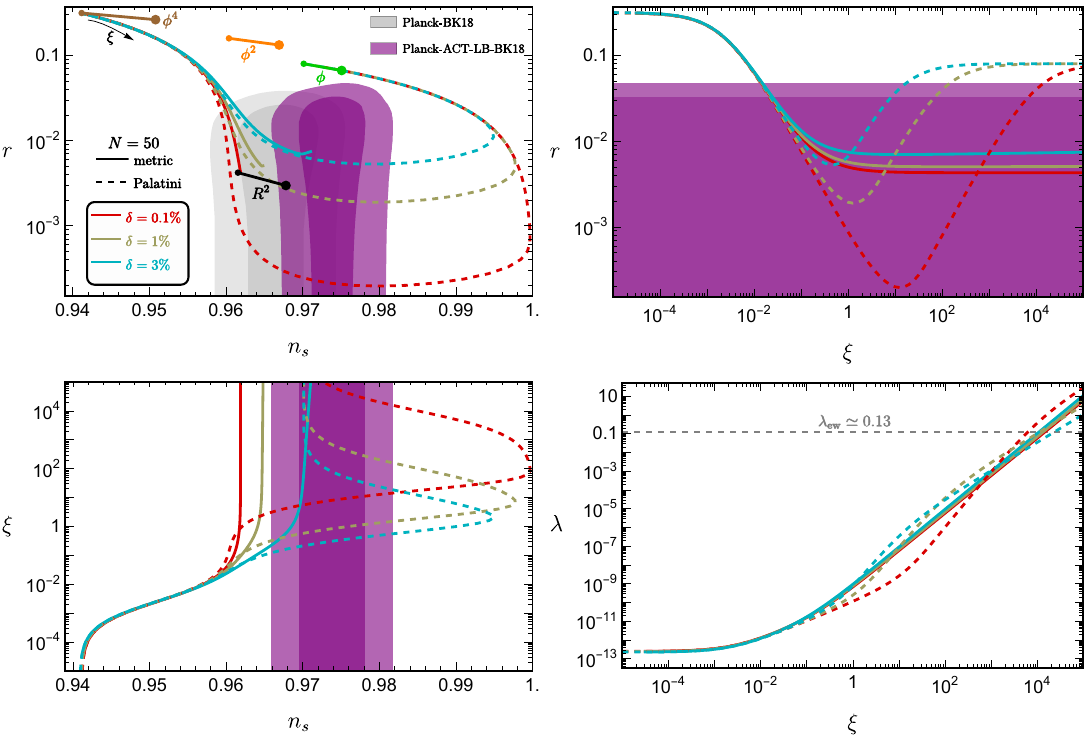}%
\caption{$r$ vs. $n_s$ (upper left panel), $r$ vs. $\xi$ (upper right),  $\xi$ vs. $n_s$ (lower left) and $\lambda$ vs. $\xi$ (lower right) for $N = 50$ $e$-folds in the metric (continuous) and Palatini formulation (dashed), with $\delta=0.1\%$, $\delta=1\%$ and $\delta=3\%$ in the loop-corrected NMC scenario. The gray (purple) areas represent the 1,2$\sigma$ allowed regions coming from the latest combination of Planck, BICEP/Keck and BAO data~\cite{BICEP:2021xfz} (from Planck, ACT, and DESI~\cite{ACT:2025tim}). For reference, we also plot the predictions of quartic (brown), quadratic (orange), linear (green) and Starobinsky~\cite{Starobinsky:1980te} (black) inflation in metric gravity, and, in the right lower panel, $\lambda_{\rm ew} \simeq 0.13$ (gray dashed line), i.e., the value of the Higgs self-quartic coupling at EW scale. The arrow in the upper left panel denotes the direction of increasing $\xi$.}
   \label{Fig:fig1}
\end{figure*}
%%%%%%%%%%%%%%%%%%%%%%%%%%%%%%%%%%%%%%%%%%%%%%%%%%%%%%%%

%%%%%%%%%%%%%%%%%%%%%%%%%%%%%%%%%%%%%%%%%%%%%%%%%%%%%%%%
\begin{figure*}[t!]
    \centering
    \includegraphics[width=\textwidth]{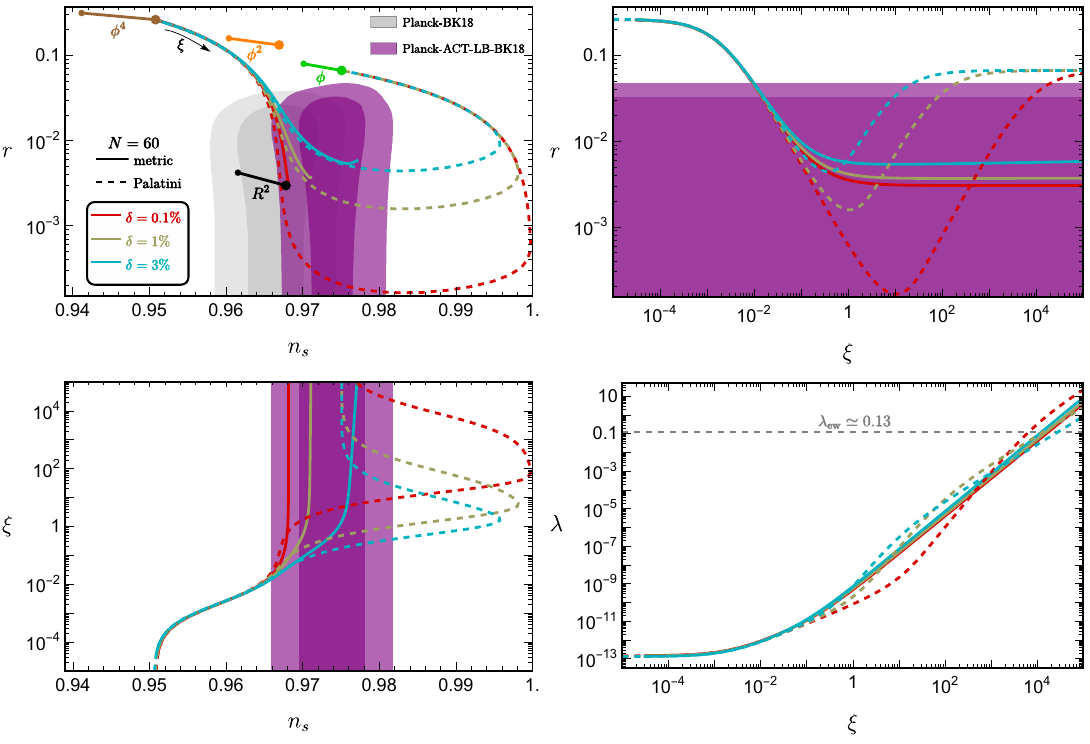}%
\caption{$r$ vs. $n_s$ (upper left panel), $r$ vs. $\xi$ (upper right),  $\xi$ vs. $n_s$ (lower left) and $\lambda$ vs. $\xi$ (lower right) for $N = 60$ $e$-folds in the metric (continuous) and Palatini formulation (dashed), with $\delta=0.1\%$, $\delta=1\%$ and $\delta=3\%$ in the loop-corrected NMC scenario. The gray (purple) areas represent the 1,2$\sigma$ allowed regions coming from the latest combination of Planck, BICEP/Keck, and BAO data~\cite{BICEP:2021xfz} (from Planck, ACT, and DESI~\cite{ACT:2025tim}). For reference, we also plot the predictions of quartic (brown), quadratic (orange), linear (green) and Starobinsky~\cite{Starobinsky:1980te} (black) inflation in metric gravity, and, in the right lower panel, $\lambda_{\rm ew} \simeq 0.13$ ( gray dashed line) i.e. the value of the Higgs self-quartic coupling at EW scale. The arrow in the upper left panel denotes the direction of increasing $\xi$.}
   \label{Fig:fig2}
\end{figure*}
%%%%%%%%%%%%%%%%%%%%%%%%%%%%%%%%%%%%%%%%%%%%%%%%%%%%%%%%
%%%%%%%%%%%%%%%%%%%%%%%%%%%%%%%%%%%%%%%%%%%%%%%%%%%%%%%%
Using Eqs.~\eqref{eq:lambdataylor} and~\eqref{eq:U}, we can write the Einstein frame scalar potential as
\be
 U(\chi)  =  \frac{\lambda \, M_P^4 \, \phi(\chi)^4}{4\left[ M_P^2 + \xi \phi(\chi)^2 \right]^2} \left[1 + \delta \ln\left(\frac{\phi(\chi)}{\MP}\right)\right]\,.
\label{eq:Ufinal}
\ee
Note that in Eq.~\eqref{eq:Ufinal} and in the following expressions, we omit the argument ``$(M_P)$'' for $\lambda$ and $\delta$ to simplify the notation.
In the slow-roll regime, the inflationary dynamics is described by the standard potential slow-roll parameters and the total number of $e$-folds that measures the duration of inflation. The potential slow-roll parameters are defined as
\be
\epsilon_U \equiv \frac{M_{\rm P}^2}{2} \left(\frac{1}{U(\chi)}\frac{{\rm d}U(\chi)}{{\rm d}\chi}\right)^2 \,, \quad
\eta_U \equiv  \frac{M_{\rm P}^2}{U(\chi)}\frac{{\rm d}^2U(\chi)}{{\rm d}\chi^2} \,,
\ee
and the number of $e$-folds are given by
\be
N = \frac{1}{M_{\rm P}^2} \int_{\chi_{\rm end}}^{\chi_\star} {\rm d}\chi\,U(\chi) \left(\frac{{\rm d}U(\chi)}{{\rm d} \chi}\right)^{-1},
\label{Ndef}
\ee
where $\chi_\star$ is the field value at the time that the pivot scale $k_\star=0.05$ Mpc$^{-1}$ left the horizon and $\chi_{\rm end}$ is the field value at the end of inflation, defined via $\epsilon_U(\chi_{\rm end}) = 1$. 
The amplitude of the scalar power spectrum is given by
\begin{equation}
 A_s = \frac{1}{24 \pi^2 M^4_P }\frac{U(\chi)}{\epsilon_U  (\chi)}\,,
\label{eq:As}
\end{equation}
and at $k_\star=0.05$ Mpc$^{-1}$ has been constrained to the value $A_s^\star \simeq 2.1 \times 10^{-9}$~\cite{Planck:2018jri}. Also, in the slow-roll approximation the tensor-to-scalar ratio $(r)$ and the spectral index of the scalar power spectrum $(n_s)$ are given by
\begin{equation}
n_s \simeq 1-6\epsilon_U +2\eta_U\,, \quad \text{and} \quad r\simeq16\epsilon_U\,, \label{eq:r:ns}
\end{equation}
respectively.

%%%%%%%%%%%%%%%%%%%%%%%%%%%%%%%%%%%%%%%%%%%%%%%%%%%%%%%%
%%%%%%%%%%%%%%%%%%%%%%%%%%%%%%%%%%%%%%%%%%%%%%%%%%%%%%%%

The corresponding numerical results are given in Figs.~\ref{Fig:fig1} and~\ref{Fig:fig2} for $N=50,60$, respectively, where we show $r$ vs.~$n_s$ (upper left panel), $r$ vs.~$\xi$ (upper right),  $\xi$ vs.~$n_s$ (lower left) and $\lambda$ vs.~$\xi$ (lower right) in the metric (continuous) and Palatini formulation (dashed), with $\delta=0.1\%$, $\delta=1\%$ and $\delta=3\%$ in the loop-corrected NMC scenario. The gray (purple) areas represent the 1,2$\sigma$ allowed regions coming from the latest combination of Planck, BICEP/Keck, and BAO data~\cite{BICEP:2021xfz} (from Planck, ACT, and DESI~\cite{ACT:2025tim}). For reference, we also plot the predictions of quartic (brown), quadratic (orange), linear (green) and Starobinsky~\cite{Starobinsky:1980te} (black) inflation in metric gravity, and, in the right lower panel, $\lambda \simeq 0.13$ (gray dashed line), i.e., the value of the Higgs self-quartic coupling at the electroweak (EW) scale. We note that the duration of reheating after inflation, and consequently the number of $e$-folds, is in general determined by the specific inflationary model (see, e.g. Refs.~\cite{Ema:2016dny, Rubio:2019ypq} for relevant studies in both the metric and Palatini formulations). As already mentioned, in this work we assume the standard benchmark values $N=50$ and $N=60$, without loss of generality.

We start by discussing the results of the metric case. As is typical of nonminimally coupled models, by increasing $\xi$, the predictions move toward %smaller (larger) values of $r$ ($n_s$). 
smaller values of $r$ and larger values of $n_s$.
When the relative loop correction $\delta$ is very small (in our case $0.1\%$), the strong coupling predictions are very close to the ones of the corresponding tree-level limit, i.e., Starobinsky inflation. By increasing $\delta$, we depart more and more from the Starobinsky limit, toward higher values of both $r$ and $n_s$. At $N=50,60$, the predictions enter the 1$\sigma$ allowed region by Ref.~\cite{ACT:2025tim} when $\delta=3\%$, while at $N=60$, this is also possible for $\delta = 1\%$. Finally, we note that this time, the strong coupling linear inflation limit~\cite{Racioppi:2018zoy} is not reached in the metric case but only in the Palatini formulation. This happens because we considered smaller values of $\delta$ with respect to the ones used in Ref.~\cite{Racioppi:2018zoy} and because in the metric case we stopped the analysis around $\xi \sim 10^4$, corresponding to $\lambda \sim 1$, which we consider a naive upper bound to ensure the perturbativity of the theory. Let us now discuss the predictions in the Palatini formulation. When $\xi$ is small, the results are indistinguishable from the ones of the metric case, until $0.01 \lesssim \xi \lesssim 0.1$, where the results of the metric and Palatini formulation start to be visibly displaced, according to the exact value of $\delta$: the smaller $\delta$, the more visible the difference. For $\xi \gtrsim 0.1$, the predictions easily enter the 1$\sigma$ allowed region of ACT~\cite{ACT:2025tim}.

We stress that in this paper, we showed only results for $\delta > 0$, naively meaning that corrections originating from bosonic degrees of freedom are dominating over the fermionic ones. The opposite case, $\delta < 0$, generates results at even lower values for $n_s$ than the tree-level ones. This has been tested numerically for both gravity formulations, but we omit the numerical results for the sake of brevity. However, it can also be proven by studying the variation of the predictions when the loop corrections are just a small correction of the tree-level ones, obtaining
\bea
 \Delta r &=&  r -  r_\text{tree-level} \approx \frac{4 \delta }{N}\,, \label{eq:delta:r} \\
 \Delta n_s &=&  n_s -  n_{s, \text{tree-level}} \approx \frac{2 \xi }{1+ 6 \varepsilon  \xi} \delta\,, \label{eq:delta:ns}
\eea
where $\varepsilon$ is the same as in Eq.~\eqref{eq:dphiE}. Such approximations are only valid in a small region where the loop-corrected results move away from the tree-level ones. However, they are enough to prove that in both formulations a positive (negative) $\delta$ implies a shift toward larger (smaller) $n_s$ values. Moreover Eq.~\eqref{eq:delta:ns} gives a naive understanding of why, at the same $\xi$ values, $\Delta n_s$ tends to larger values in the Palatini case. 
The details about the derivation of Eqs.~\eqref{eq:delta:r} and~\eqref{eq:delta:ns} are provided in Appendix~\ref{Appendix}. Moreover, see also Refs.~\cite{Barvinsky:2008ia, Barvinsky:2009ii}.
\begin{figure}[t!]
    \centering
    \includegraphics[width=0.485\textwidth]{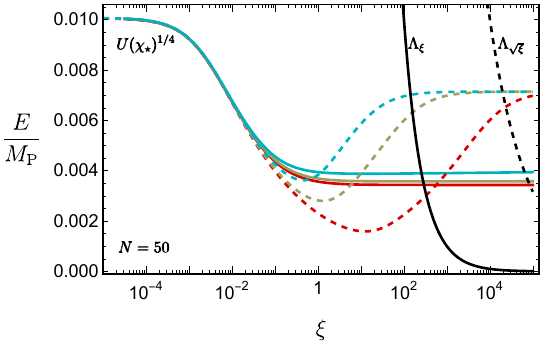}%
\caption{The value of the potential~\eqref{eq:Ufinal} at horizon crossing for $N=50$ as a function of $\xi$ for the metric and Palatini formulations (colors as in Figs.~\ref{Fig:fig1} and~\ref{Fig:fig2}). The cutoff scales $\Lambda_\xi$ (solid black line) and $\Lambda_{\sqrt{\xi}}$ (dashed black line) are also shown to indicate the region of validity of our results.}
   \label{Fig:fig3}
\end{figure}

It is worthwhile to briefly address the issue of perturbative unitarity violation  (at the scale $ \Lambda_\xi = M_P / \xi $ in metric gravity~\cite{Burgess:2009ea, Barbon:2009ya} and $ \Lambda_{\sqrt\xi} = M_P / \sqrt\xi $ in Palatini gravity~\cite{Bauer:2010jg}), which is a common concern in nonminimally coupled models. In standard Higgs-like inflation the unitarity issue can be avoided in both gravity formulations~\cite{Bauer:2010jg} by allowing $\lambda$ to be a free parameter. The addition of the radiative correction does not spoil this feature. In our analysis, we ignored the effect of operators suppressed by tree-level unitarity violation scales (e.g., Ref.~\cite{Bezrukov:2010jz}), which might affect the validity of the computations of radiative corrections and/or the inflationary physics (e.g., Refs.~\cite{Branchina:2013jra, Bezrukov:2014ipa, Bezrukov:2014bra}). 
However, in both gravity formulations, it is possible to enter the ACT 2$\sigma$ allowed region for $ \xi \lesssim 1$, implying that $\Lambda$ is pushed well above the inflationary scale\footnote{Note that the constraint $A_s^\star \simeq 2.1 \times 10^{-9}$~\cite{Planck:2018jri} combined with $r \lesssim 0.1$ ensures a sub-Planckian inflationary scale.} $U(\chi_*)^{1/4}$ (confirming the validity of our assumptions in the computations of the radiative corrections) and even above the Planck scale, where an appropriate UV completion of our theory is inevitably required. Nevertheless, it is interesting to check for which $\xi>1$ values, the unitarity violation scale remains above the inflationary one. In Fig.~\ref{Fig:fig3} we plot $U(\chi_*)^{1/4}$ for $N=50$ as a function of $\xi$ for the metric and Palatini formulations (colors as in Figs.~\ref{Fig:fig1} and~\ref{Fig:fig2}). The corresponding cutoff scales $\Lambda_\xi$ (solid black line) and $\Lambda_{\sqrt{\xi}}$ (dashed black line) are also depicted. We omit a plot for $N=60$, because it is very similar to Fig.~\ref{Fig:fig3} and therefore does not carry any additional relevant information. From Fig.~\ref{Fig:fig3} we can see that unitarity can be preserved also for $\xi > 1$. In particular, this is possible respectively for $\xi \lesssim 300$ in the metric formulation and $\xi \lesssim 2 \times 10^4$ in the Palatini one.
%We note that the obtained results mitigate such an issue. In both gravity formulations, it is possible to enter the ACT 2$\sigma$ allowed region for $ \xi \lesssim 1$, implying that $\Lambda$ is pushed well above the inflationary regime (confirming the validity of our assumptions in the computations of the radi\cite{Bauer:2010jg} bye corrections) and even above the Planck scale, where an appropriate UV completion of our theory is inevitably required.

We stress that in this analysis we have been agnostic about the details of the underlying UV theory and the nature of the inflaton scalar. However,  it might be possible to identify the inflaton with the SM Higgs boson, with a couple of additional tweaks.
In the conventional SM Higgs-inflation scenario, it is quite common to work in a setup where both the Higgs self-quartic coupling and the corresponding beta function are $\lambda_H^\text{SM} \simeq \beta_H^\text{SM} \simeq 0$ at $M_P$ (e.g., Refs.~\cite{Bezrukov:2014bra, Shaposhnikov:2020fdv} and references therein). Therefore, the radiative corrections are typically expected to be much smaller than $1\%$. However, in such a regime, since $\beta_H^\text{SM}=0$, $\lambda_H$ is extremely sensitive to any contribution coming from beyond the Standard Model (BSM) physics. Therefore, the radiative corrections can become significant in scenarios where the running of the Higgs is modified by BSM physics, for instance, in order to solve the vacuum stability problem without adjusting the top-Yukawa coupling (e.g.~\cite{Kadastik:2011aa, Gabrielli:2013hma, Okada:2015lia}). Therefore, our results show how and in what amount BSM physics needs to intervene at high scale so that the expansion in Eq.~\eqref{eq:lrun} is still a viable approximation. In the common scenarios where $\beta_H^\text{SM} \simeq 0$, the validity of Eq.~\eqref{eq:lrun} would then be all due to BSM physics.

To conclude, we remark that in the region where the parameter spaces are compatible, our results agree with the previous results of Refs.~\cite{Barvinsky:2008ia, Barvinsky:2009ii, Okada:2010jf, Okada:2010jd, Okada:2011en, Okada:2015lia}.

\section{Conclusions}
\label{conclusions}

In this work, we have shown that by incorporating a running self-coupling for the inflaton scalar, the model can naturally accommodate larger values of $n_s$, bringing it into excellent agreement with the $1\sigma$ region of the ACT data. This effect is even more pronounced in the Palatini formulation of the theory. As known, keeping the self-quartic coupling $\lambda$ as a free parameter, the required values of the nonminimal coupling $\xi$ allow one to push the issue of perturbative unitarity above the inflationary scale. Such a feature remains also in presence of radiative corrections. These findings suggest that quantum corrections play a crucial role in reconciling NMC inflation with the latest observational data and motivate further exploration of radiative effects and their implications for inflationary predictions.

\section*{Acknowledgements}
This work was supported by the Estonian Research Council grants MOB3JD1202, PSG761, PRG1055, RVTT3, RVTT7, and the CoE program TK202 ``Foundations of the Universe'’. This article is based upon work from COST Action CosmoVerse CA21136, supported by COST (European Cooperation in Science and Technology).

\appendix

\section{Details on the treatment of radiative corrections} \label{Appendix:CW}
In agreement with our definition of Higgs-like we focus on a 1-loop effective quartic scalar potential, 
\begin{equation}
  V_\text{eff}(\phi,\mu) = \frac{\lambda_\text{eff}(\phi,\mu)}{4} \phi^4\,,
  \label{eq:Veff_2}
\end{equation}
where $\mu$ is the renormalization scale and $\lambda_\text{eff}(\phi,\mu)$ is the  effective quartic coupling
\be
\lambda_\text{eff}(\phi,\mu) 
%&=& \lambda_{\rm tree} + \lambda_{\rm 1-loop}(\phi)+\cdots \nn\\
= \lambda_\text{run}(\mu) + \lambda_\text{CW}(\phi,\mu)\,,
\ee
containing both the running contribution $\lambda_\text{run}(\mu)$ and the Colewan-Weinberg \cite{Coleman:1973jx} (CW)  one $\lambda_\text{CW}(\phi,\mu)$. Such corrections to $\lambda$ originate from the full particle spectrum of a UV-complete theory. However, as mentioned before, we remain agnostic about the exact particle content of the theory (apart from the assumptions of only classically scale-invariant interactions in the Jordan frame) in order to provide a model-independent study.
 
The running of $\lambda$ is governed by its $\beta$-function, $\beta_\lambda (\mu) = \dd\lambda/\dd\ln \mu $, where $\mu$ is the renormalization scale.
The exact expression of $\beta$ depends on the full UV-completed theory. However, setting aside those details, we can expand the running quartic coupling as a Taylor series:
\begin{equation}
\lambda_\text{run}(\mu) = \bar\lambda(\mu_0) + \sum_{n=1}^\infty \frac{\beta_{n-1}}{n!} \ln^n \left(\frac{\mu}{\mu_0}\right) \,, 
\label{eq:lambdataylor}
\end{equation}
where $\mu_0$ is a reference scale needed to set the boundary condition of the renormalization group equation, and $\beta_n$ denotes the $n$-th derivative of the $\beta$-function at this scale. Assuming that only the leading-order term of the expansion is relevant during inflation we obtain
\begin{equation}
  \lambda_\text{run}(\mu) \simeq \bar\lambda(\mu_0)  + \beta_0 \ln\left(\frac{\mu}{\mu_0}\right)\,,
  \label{eq:lrun:app}
\end{equation}
where $\beta_0 = \beta(\mu_0)$. Let us move to the CW part. Following Ref.~\cite{Gildener:1976ih}, we can write the CW contribution at 1-loop as
\be
\lambda_\text{CW}(\phi,\mu)	=  B \ln\left(\frac{\kappa \phi}{\mu}\right)= A + B \ln\left(\frac{\phi}{\mu}\right)\,,
\label{eq:lambdaCW}
\ee
where $A= B \ln \kappa$ and again we used the assumption of $\phi$ possessing only classically scale-invariant interactions in the Jordan frame with the UV-complete particle spectrum, as assumed before in our definition of Higgs-like. The constant $\kappa$ contains information about the couplings of $\phi$ to the other particles in the theory.

In~\eqref{eq:lambdaCW}, we kept the leading-order approximation as well; otherwise, $A$ and $B$  would be functions of $\mu$. We stress that $\kappa$ is instead not a function of $\mu$, because considering a $\mu$-dependence on $\kappa$ would have been a 2-loop and not a 1-loop correction (e.g. Ref.~\cite{Neubert:2019mrz} and references therein).
As we remain agnostic about the exact particle content, we will also treat $A$ and $B$ as free parameters. 
Using Eqs.~\eqref{eq:lrun:app} and~\eqref{eq:lambdaCW} we obtain
\be
\lambda_\text{eff}(\phi,\mu) \simeq  \bar\lambda(\mu_0)  + A + \beta_0 \ln\left(\frac{\mu}{\mu_0}\right) +  B \ln\left(\frac{\phi}{\mu}\right)\,.
\label{eq:lambdaeffnew}
\ee
It is now easy to check that the consistency of the Callan-Symanzik equation, $\frac{\text{d}}{\text{d}\mu} V_\text{eff}=0$, \cite{Callan:1970, Symanzik:1970} implies $\beta_0 = B$, so that 
\be
\lambda_\text{eff}(\phi,\mu) \simeq  \lambda(\mu_0)  + \beta_0 \ln\left(\frac{\phi}{\mu_0}\right)\,,
\label{eq:lambdarun}
\ee
where 
\be
\lambda(\mu_0)= \bar\lambda(\mu_0) + A\,, \label{eq:newlambda}
\ee
and any explicit dependence on $\mu$ is canceled, as long as the approximations in Eqs.~\eqref{eq:lrun:app} and~\eqref{eq:lambdaCW} hold (for some explicit realization, see, e.g., Refs.~\cite{Kannike:2014mia,Racioppi:2019jsp}). Additionally, the choice of $\mu_0$ carries no physical significance as long as $\lambda(\mu_0)$ and $\beta_0$ are adjusted accordingly within the region where Eq.~\eqref{eq:lambdarun} is valid. Thus, for numerical convenience, we set the reference scale to $\mu_0 = M_P$ (e.g., Refs.~\cite{Racioppi:2018zoy, Racioppi:2019jsp} and references therein). Ultimately, it is convenient to define the relative one-loop correction as $\delta = \beta_0 / \lambda$, treat it as a free parameter and obtain
\begin{equation}
  \lambda_{\rm eff}(\phi) \simeq \lambda(M_P) \left[1 + \delta(M_P) \ln\left(\frac{\phi}{M_P}\right)\right]\,,
\end{equation}
which exactly the same as $\lambda_{\rm eff}(\phi)$ given in Eq.~\eqref{eq:lambdarun}.

To conclude this Appendix, we comment on an alternative method, useful especially when it is not possible to explicitly cancel the dependence on the renormalization scale $\mu$, which is to choose a convenient value $\mu \approx c \phi$, where $c$ can change according to the problem at hand, leading to 
\be
\lambda_\text{eff}(\phi,\mu) = \lambda_\text{run}\left(c \phi \right)  +  B (c \phi) \ln \left(\frac{\kappa}{c}\right)\,,
\label{eq:lambdaeffmu}
\ee
where we have restored the dependence of $B$ on $\mu$. Choosing $c = \kappa$ (e.g., Refs.~\cite{DeSimone:2008ei, Barvinsky:2008ia, Barvinsky:2009ii} and references therein) would exactly cancel the shift and move all the $\phi$-dependence inside $\lambda_\text{run}$. Another popular choice is $c=1$ (e.g., Refs.~\cite{Espinosa:2015qea, Kannike:2015apa} and references therein). Then, in case the approximations~\eqref{eq:lrun:app} and \eqref{eq:lambdaCW} hold, we get the same as~\eqref{eq:lambdarun}, but with
\be
\lambda(\mu_0) =  \bar\lambda(\mu_0)  + A +  (\beta_0-B) \ln c\,,
\label{eq:lambda:other}
\ee
which can be used again as the effective free parameter. Note that~\eqref{eq:lambda:other} reduces to~\eqref{eq:newlambda} when $\beta_0 = B$, in perfect agreement with the consistency of the Callan-Symanzik equation.

\section{Details on the computations of the inflationary observables} \label{Appendix}
In this Appendix, we provide the details of the computations that lead to the approximated results given in Eqs.~\eqref{eq:delta:r} and~\eqref{eq:delta:ns}. As we said, we look for results nearby the tree-level ones; therefore, the first step is to compute the number of $e$-folds in the case 
\bea
\phi_\star & \simeq & \phi_{\star, \text{ tree-level}}+ \Delta \phi_\star \label{eq:phiN:app} \, ,\\
\phi_{\star, \text{ tree-level}} & \simeq & M_P \frac{2 \sqrt{2 N}}{\sqrt{1+ 6 \varepsilon  \xi}} \, , \label{eq:phiN:tree}
\eea
where $\phi_{\star, \text{ tree-level}}$ is the tree-level result in the approximations of big $N$ and $\sqrt\xi\phi \gg M_P$~\cite{Bauer:2008zj}, while $\Delta \phi_\star$ is the correction to be evaluated in terms of $\delta$. Computing the number of $e$-folds~\eqref{Ndef} using Eqs.~\eqref{eq:dphiE}, \eqref{eq:phiN:app}, and~\eqref{eq:phiN:tree}, keeping the aforementioned approximations and the leading order in $\delta$ and $\Delta \phi_\star$, we obtain
\bea
 N &=&  \frac{1}{M_{\rm P}^2} \int_{\phi_{\rm end}}^{\phi_\star} {\rm d}\phi \left(\frac{{\rm d}\chi}{{\rm d}\phi}\right)^2 \,U(\phi) \left(\frac{{\rm d}U(\phi)}{{\rm d} \phi}\right)^{-1} \, , \nonumber \\
  &\simeq & N + 2   N \Delta \phi_\star-\frac{ N (4 \xi N +1)}{4 ( 1 + 6 \varepsilon\xi)} \delta \, ,
\eea
which implies
\be
 \Delta \phi_\star \simeq \frac{ (4 \xi N +1)}{8 ( 1 + 6 \varepsilon\xi)} \delta \, .
 \ee
Computing then Eqs.~\eqref{eq:r:ns} at the leading order in $\delta$, we obtain
\bea
 r & \simeq & r_\text{tree-level}  +\frac{4 \delta }{N}\, , \label{eq:r:loop:app}\\
 n_s &\simeq &  n_{s, \text{tree-level}} + \frac{2 \delta  \xi }{1+6 \varepsilon\xi}\, , \label{eq:ns:loop:app}
\eea
where the tree-level results are~\cite{Bezrukov:2007ep,Bauer:2008zj}
\bea
 r_\text{tree-level} & \simeq & \frac{12 \varepsilon}{N^2} + \frac{2}{N^2 \xi } \, , \label{eq:r:tree}\\
 n_{s, \text{tree-level}} &\simeq & \, 1-\frac{2}{N}  \, . \label{eq:ns:tree}
\eea
It is easy to check from Eqs.~\eqref{eq:r:loop:app} and~\eqref{eq:ns:loop:app} that we obtain the results of Eqs.~\eqref{eq:delta:r} and~\eqref{eq:delta:ns}. 

\bibliography{references}

@article{ACT:2025tim,
    author = "Calabrese, Erminia and others",
    collaboration = "ACT",
    title = "{The Atacama Cosmology Telescope: DR6 Constraints on Extended Cosmological Models}",
    eprint = "2503.14454",
    archivePrefix = "arXiv",
    primaryClass = "astro-ph.CO",
    reportNumber = "FERMILAB-PUB-25-0157-PPD",
    month = "3",
    year = "2025"
}

@article{Racioppi:2018zoy,
    author = "Racioppi, Antonio",
    title = "{New universal attractor in nonminimally coupled gravity: Linear inflation}",
    eprint = "1801.08810",
    archivePrefix = "arXiv",
    primaryClass = "astro-ph.CO",
    doi = "10.1103/PhysRevD.97.123514",
    journal = "Phys. Rev. D",
    volume = "97",
    number = "12",
    pages = "123514",
    year = "2018"
}

@article{Racioppi:2019jsp,
    author = "Racioppi, Antonio",
    title = "{Non-Minimal (Self-)Running Inflation: Metric vs. Palatini Formulation}",
    eprint = "1912.10038",
    archivePrefix = "arXiv",
    primaryClass = "hep-ph",
    doi = "10.1007/JHEP01(2021)011",
    journal = "JHEP",
    volume = "21",
    pages = "011",
    year = "2020"
}

@article{Kallosh:2025rni,
    author = "Kallosh, Renata and Linde, Andrei and Roest, Diederik",
    title = "{A simple scenario for the last ACT}",
    eprint = "2503.21030",
    archivePrefix = "arXiv",
    primaryClass = "hep-th",
    month = "3",
    year = "2025"
}

@article{Marzola:2016xgb,
      author         = "Marzola, Luca and Racioppi, Antonio",
      title          = "{Minimal but non-minimal inflation and electroweak
                        symmetry breaking}",
      journal        = "JCAP",
      volume         = "1610",
      year           = "2016",
      number         = "10",
      pages          = "010",
      doi            = "10.1088/1475-7516/2016/10/010",
      eprint         = "1606.06887",
      archivePrefix  = "arXiv",
      primaryClass   = "hep-ph",
      SLACcitation   = "%%CITATION = ARXIV:1606.06887;%%"
}

@article{Kannike:2015apa,
      author         = "Kannike, Kristjan and H{\"u}tsi, Gert and Pizza, Liberato
                        and Racioppi, Antonio and Raidal, Martti and Salvio,
                        Alberto and Strumia, Alessandro",
      title          = "{Dynamically Induced Planck Scale and Inflation}",
      journal        = "JHEP",
      volume         = "05",
      year           = "2015",
      pages          = "065",
      doi            = "10.1007/JHEP05(2015)065",
      eprint         = "1502.01334",
      archivePrefix  = "arXiv",
      primaryClass   = "astro-ph.CO",
      reportNumber   = "IFUP-TH-2015, IFT-UAM-CSIC-15-015,
                        IFUP-TH-2015-AND-IFT-UAM-CSIC-15-015",
      SLACcitation   = "%%CITATION = ARXIV:1502.01334;%%"
}

@article{Kannike:2014mia,
      author         = "Kannike, Kristjan and Racioppi, Antonio and Raidal,
                        Martti",
      title          = "{Embedding inflation into the Standard Model - more
                        evidence for classical scale invariance}",
      journal        = "JHEP",
      volume         = "06",
      year           = "2014",
      pages          = "154",
      doi            = "10.1007/JHEP06(2014)154",
      eprint         = "1405.3987",
      archivePrefix  = "arXiv",
      primaryClass   = "hep-ph",
      SLACcitation   = "%%CITATION = ARXIV:1405.3987;%%"
}

@article{Gabrielli:2013hma,
      author         = "Gabrielli, Emidio and Heikinheimo, Matti and Kannike,
                        Kristjan and Racioppi, Antonio and Raidal, Martti and
                        Spethmann, Christian",
      title          = "{Towards Completing the Standard Model: Vacuum Stability,
                        EWSB and Dark Matter}",
      journal        = "Phys. Rev.",
      volume         = "D89",
      year           = "2014",
      number         = "1",
      pages          = "015017",
      doi            = "10.1103/PhysRevD.89.015017",
      eprint         = "1309.6632",
      archivePrefix  = "arXiv",
      primaryClass   = "hep-ph",
      SLACcitation   = "%%CITATION = ARXIV:1309.6632;%%"
}

@article{Bezrukov:2007ep,
      author         = "Bezrukov, Fedor L. and Shaposhnikov, Mikhail",
      title          = "{The Standard Model Higgs boson as the inflaton}",
      journal        = "Phys. Lett.",
      volume         = "B659",
      year           = "2008",
      pages          = "703-706",
      doi            = "10.1016/j.physletb.2007.11.072",
      eprint         = "0710.3755",
      archivePrefix  = "arXiv",
      primaryClass   = "hep-th",
      SLACcitation   = "%%CITATION = ARXIV:0710.3755;%%"
}

@article{Koivisto:2005yc,
      author         = "Koivisto, Tomi and Kurki-Suonio, Hannu",
      title          = "{Cosmological perturbations in the palatini formulation
                        of modified gravity}",
      journal        = "Class. Quant. Grav.",
      volume         = "23",
      year           = "2006",
      pages          = "2355-2369",
      doi            = "10.1088/0264-9381/23/7/009",
      eprint         = "astro-ph/0509422",
      archivePrefix  = "arXiv",
      primaryClass   = "astro-ph",
      reportNumber   = "HIP-2005-38-TH",
      SLACcitation   = "%%CITATION = ASTRO-PH/0509422;%%"
}

@article{Bauer:2010jg,
      author         = "Bauer, Florian and Demir, Durmus A.",
      title          = "{Higgs-Palatini Inflation and Unitarity}",
      journal        = "Phys. Lett.",
      volume         = "B698",
      year           = "2011",
      pages          = "425-429",
      doi            = "10.1016/j.physletb.2011.03.042",
      eprint         = "1012.2900",
      archivePrefix  = "arXiv",
      primaryClass   = "hep-ph",
      reportNumber   = "IZTECH-P-10-06, ICCUB-10-200",
      SLACcitation   = "%%CITATION = ARXIV:1012.2900;%%"
}

@article{Barbon:2009ya,
author = "Barbon, J. L. F. and Espinosa, J. R.",
title = "{On the Naturalness of Higgs Inflation}",
eprint = "0903.0355",
archivePrefix = "arXiv",
primaryClass = "hep-ph",
reportNumber = "IFT-UAM-CSIC-09-10, UAB-FT-665",
doi = "10.1103/PhysRevD.79.081302",
journal = "Phys. Rev. D",
volume = "79",
pages = "081302",
year = "2009"
}

@article{Starobinsky:1980te,
      author         = "Starobinsky, Alexei A.",
      title          = "{A New Type of Isotropic Cosmological Models Without
                        Singularity}",
      journal        = "Phys. Lett.",
      volume         = "B91",
      year           = "1980",
      pages          = "99-102",
      doi            = "10.1016/0370-2693(80)90670-X",
      SLACcitation   = "%%CITATION = PHLTA,B91,99;%%"
}

@article{Linde:1981mu,
	Author = {Linde, Andrei D.},
	Doi = {10.1016/0370-2693(82)91219-9},
	Journal = {Phys.Lett.},
	Pages = {389-393},
	Reportnumber = {LEBEDEV-81-229},
	Slaccitation = {%%CITATION = PHLTA,B108,389;%%},
	Title = {{A New Inflationary Universe Scenario: A Possible Solution of the Horizon, Flatness, Homogeneity, Isotropy and Primordial Monopole Problems}},
	Volume = {B108},
	Year = {1982}}

@article{Guth:1980zm,
	Author = {Guth, Alan H.},
	Doi = {10.1103/PhysRevD.23.347},
	Journal = {Phys.Rev.},
	Pages = {347-356},
	Reportnumber = {SLAC-PUB-2576},
	Slaccitation = {%%CITATION = PHRVA,D23,347;%%},
	Title = {{The Inflationary Universe: A Possible Solution to the Horizon and Flatness Problems}},
	Volume = {D23},
	Year = {1981}}

@article{Bauer:2008zj,
      author         = "Bauer, Florian and Demir, Durmus A.",
      title          = "{Inflation with Non-Minimal Coupling: Metric versus
                        Palatini Formulations}",
      journal        = "Phys. Lett.",
      volume         = "B665",
      year           = "2008",
      pages          = "222-226",
      doi            = "10.1016/j.physletb.2008.06.014",
      eprint         = "0803.2664",
      archivePrefix  = "arXiv",
      primaryClass   = "hep-ph",
      reportNumber   = "DESY-08-033, IZTECH-P-08-02",
      SLACcitation   = "%%CITATION = ARXIV:0803.2664;%%"
}

@article{Espinosa:2015qea,
      author         = "Espinosa, Jose R. and Giudice, Gian F. and Morgante,
                        Enrico and Riotto, Antonio and Senatore, Leonardo and
                        Strumia, Alessandro and Tetradis, Nikolaos",
      title          = "{The cosmological Higgstory of the vacuum instability}",
      journal        = "JHEP",
      volume         = "09",
      year           = "2015",
      pages          = "174",
      doi            = "10.1007/JHEP09(2015)174",
      eprint         = "1505.04825",
      archivePrefix  = "arXiv",
      primaryClass   = "hep-ph",
      reportNumber   = "CERN-PH-TH-2015-119, IFUP-TH-2015",
      SLACcitation   = "%%CITATION = ARXIV:1505.04825;%%"
}

@article{Rasanen:2017ivk,
      author         = "Rasanen, Syksy and Wahlman, Pyry",
      title          = "{Higgs inflation with loop corrections in the Palatini
                        formulation}",
      journal        = "JCAP",
      volume         = "1711",
      year           = "2017",
      number         = "11",
      pages          = "047",
      doi            = "10.1088/1475-7516/2017/11/047",
      eprint         = "1709.07853",
      archivePrefix  = "arXiv",
      primaryClass   = "astro-ph.CO",
      reportNumber   = "HIP-2017-23-TH, KOBE-COSMO-17-12",
      SLACcitation   = "%%CITATION = ARXIV:1709.07853;%%"
}

@article{George:2015nza,
	Archiveprefix = {arXiv},
	Author = {George, Damien P. and Mooij, Sander and Postma, Marieke},
	Doi = {10.1088/1475-7516/2016/04/006},
	Eprint = {1508.04660},
	Journal = {JCAP},
	Number = {04},
	Pages = {006},
	Primaryclass = {hep-th},
	Slaccitation = {%%CITATION = ARXIV:1508.04660;%%},
	Title = {{Quantum corrections in Higgs inflation: the Standard Model case}},
	Volume = {1604},
	Year = {2016}}

@article{Barvinsky:2008ia,
      author         = "Barvinsky, A. O. and 
      shchik, A. {\relax Yu}. and
                        Starobinsky, A. A.",
      title          = "{Inflation scenario via the Standard Model Higgs boson
                        and LHC}",
      journal        = "JCAP",
      volume         = "0811",
      year           = "2008",
      pages          = "021",
      doi            = "10.1088/1475-7516/2008/11/021",
      eprint         = "0809.2104",
      archivePrefix  = "arXiv",
      primaryClass   = "hep-ph",
      SLACcitation   = "%%CITATION = ARXIV:0809.2104;%%"
}

@article{DeSimone:2008ei,
      author         = "De Simone, Andrea and Hertzberg, Mark P. and Wilczek,
                        Frank",
      title          = "{Running Inflation in the Standard Model}",
      journal        = "Phys. Lett.",
      volume         = "B678",
      year           = "2009",
      pages          = "1-8",
      doi            = "10.1016/j.physletb.2009.05.054",
      eprint         = "0812.4946",
      archivePrefix  = "arXiv",
      primaryClass   = "hep-ph",
      reportNumber   = "MIT-CTP-4008",
      SLACcitation   = "%%CITATION = ARXIV:0812.4946;%%"
}

@article{Fumagalli:2016lls,
      author         = "Fumagalli, Jacopo and Postma, Marieke",
      title          = "{UV (in)sensitivity of Higgs inflation}",
      journal        = "JHEP",
      volume         = "05",
      year           = "2016",
      pages          = "049",
      doi            = "10.1007/JHEP05(2016)049",
      eprint         = "1602.07234",
      archivePrefix  = "arXiv",
      primaryClass   = "hep-ph",
      SLACcitation   = "%%CITATION = ARXIV:1602.07234;%%"
}

@article{Bezrukov:2017dyv,
      author         = "Bezrukov, Fedor and Pauly, Martin and Rubio, Javier",
      title          = "{On the robustness of the primordial power spectrum in
                        renormalized Higgs inflation}",
      year           = "2017",
      eprint         = "1706.05007",
      archivePrefix  = "arXiv",
      primaryClass   = "hep-ph",
      SLACcitation   = "%%CITATION = ARXIV:1706.05007;%%"
}

@article{Aoki:2025wld,
    author = "Aoki, Shuntaro and Otsuka, Hajime and Yanagita, Ryota",
    title = "{Higgs-Modular Inflation}",
    eprint = "2504.01622",
    archivePrefix = "arXiv",
    primaryClass = "hep-ph",
    reportNumber = "RIKEN-iTHEMS-Report-25, KYUSHU-HET-317",
    month = "4",
    year = "2025"
}

@article{ACT:2025fju,
    author = "Louis, Thibaut and others",
    collaboration = "ACT",
    title = "{The Atacama Cosmology Telescope: DR6 Power Spectra, Likelihoods and $\Lambda$CDM Parameters}",
    eprint = "2503.14452",
    archivePrefix = "arXiv",
    primaryClass = "astro-ph.CO",
    reportNumber = "FERMILAB-PUB-25-0071-PPD",
    month = "3",
    year = "2025"
}

@article{BICEP:2021xfz,
    author = "Ade, P. A. R. and others",
    collaboration = "BICEP, Keck",
    title = "{Improved Constraints on Primordial Gravitational Waves using Planck, WMAP, and BICEP/Keck Observations through the 2018 Observing Season}",
    eprint = "2110.00483",
    archivePrefix = "arXiv",
    primaryClass = "astro-ph.CO",
    doi = "10.1103/PhysRevLett.127.151301",
    journal = "Phys. Rev. Lett.",
    volume = "127",
    number = "15",
    pages = "151301",
    year = "2021"
}

@article{DESI:2024mwx,
    author = "Adame, A. G. and others",
    collaboration = "DESI",
    title = "{DESI 2024 VI: cosmological constraints from the measurements of baryon acoustic oscillations}",
    eprint = "2404.03002",
    archivePrefix = "arXiv",
    primaryClass = "astro-ph.CO",
    reportNumber = "FERMILAB-PUB-24-0154-PPD",
    doi = "10.1088/1475-7516/2025/02/021",
    journal = "JCAP",
    volume = "02",
    pages = "021",
    year = "2025"
}

@article{Planck:2018jri,
    author = "Akrami, Y. and others",
    collaboration = "Planck",
    title = "{Planck 2018 results. X. Constraints on inflation}",
    eprint = "1807.06211",
    archivePrefix = "arXiv",
    primaryClass = "astro-ph.CO",
    doi = "10.1051/0004-6361/201833887",
    journal = "Astron. Astrophys.",
    volume = "641",
    pages = "A10",
    year = "2020"
}

@article{Sato:1981qmu,
    author = "Sato, Katsuhiko",
    title = "{First-order phase transition of a vacuum and the expansion of the Universe}",
    doi = "10.1093/mnras/195.3.467",
    journal = "Mon. Not. Roy. Astron. Soc.",
    volume = "195",
    number = "3",
    pages = "467--479",
    year = "1981"
}

@article{Kazanas:1980tx,
    author = "Kazanas, D.",
    title = "{Dynamics of the Universe and Spontaneous Symmetry Breaking}",
    doi = "10.1086/183361",
    journal = "Astrophys. J. Lett.",
    volume = "241",
    pages = "L59--L63",
    year = "1980"
}

@article{Gialamas:2023flv,
    author = "Gialamas, Ioannis D. and Karam, Alexandros and Pappas, Thomas D. and Tomberg, Eemeli",
    title = "{Implications of Palatini gravity for inflation and beyond}",
    eprint = "2303.14148",
    archivePrefix = "arXiv",
    primaryClass = "gr-qc",
    doi = "10.1142/S0219887823300076",
    journal = "Int. J. Geom. Meth. Mod. Phys.",
    volume = "20",
    number = "13",
    pages = "2330007",
    year = "2023"
}

@article{Berera:2025vsu,
    author = "Berera, Arjun and Brahma, Suddhasattwa and Qiu, Zizang and O. Ramos, Rudnei and Rodrigues, Gabriel S.",
    title = "{The early universe is $\textit{ACT}$-ing $\textit{warm}$}",
    eprint = "2504.02655",
    archivePrefix = "arXiv",
    primaryClass = "hep-th",
    month = "4",
    year = "2025"
}

@article{Dioguardi:2025vci,
    author = "Dioguardi, Christian and Iovino, Antonio J. and Racioppi, Antonio",
    title = "{Fractional attractors in light of the latest ACT observations}",
    eprint = "2504.02809",
    archivePrefix = "arXiv",
    primaryClass = "gr-qc",
    month = "4",
    year = "2025"
}

@article{Brahma:2025dio,
    author = "Brahma, Suddhasattwa and Calder\'on-Figueroa, Jaime",
    title = "{Is the CMB revealing signs of pre-inflationary physics?}",
    eprint = "2504.02746",
    archivePrefix = "arXiv",
    primaryClass = "astro-ph.CO",
    month = "4",
    year = "2025"
}

@article{Shaposhnikov:2020fdv,
    author = "Shaposhnikov, Mikhail and Shkerin, Andrey and Zell, Sebastian",
    title = "{Quantum Effects in Palatini Higgs Inflation}",
    eprint = "2002.07105",
    archivePrefix = "arXiv",
    primaryClass = "hep-ph",
    doi = "10.1088/1475-7516/2020/07/064",
    journal = "JCAP",
    volume = "07",
    pages = "064",
    year = "2020"
}

@article{Ema:2016dny,
    author = "Ema, Yohei and Jinno, Ryusuke and Mukaida, Kyohei and Nakayama, Kazunori",
    title = "{Violent Preheating in Inflation with Nonminimal Coupling}",
    eprint = "1609.05209",
    archivePrefix = "arXiv",
    primaryClass = "hep-ph",
    doi = "10.1088/1475-7516/2017/02/045",
    journal = "JCAP",
    volume = "02",
    pages = "045",
    year = "2017"
}

@article{Rubio:2019ypq,
    author = "Rubio, Javier and Tomberg, Eemeli S.",
    title = "{Preheating in Palatini Higgs inflation}",
    eprint = "1902.10148",
    archivePrefix = "arXiv",
    primaryClass = "hep-ph",
    doi = "10.1088/1475-7516/2019/04/021",
    journal = "JCAP",
    volume = "04",
    pages = "021",
    year = "2019"
}

@article{Bezrukov:2014bra,
    author = "Bezrukov, Fedor and Shaposhnikov, Mikhail",
    title = "{Higgs inflation at the critical point}",
    eprint = "1403.6078",
    archivePrefix = "arXiv",
    primaryClass = "hep-ph",
    reportNumber = "CERN-PH-TH-2014-082",
    doi = "10.1016/j.physletb.2014.05.074",
    journal = "Phys. Lett. B",
    volume = "734",
    pages = "249--254",
    year = "2014"
}

@article{Bezrukov:2014ipa,
    author = "Bezrukov, Fedor and Rubio, Javier and Shaposhnikov, Mikhail",
    title = "{Living beyond the edge: Higgs inflation and vacuum metastability}",
    eprint = "1412.3811",
    archivePrefix = "arXiv",
    primaryClass = "hep-ph",
    doi = "10.1103/PhysRevD.92.083512",
    journal = "Phys. Rev. D",
    volume = "92",
    number = "8",
    pages = "083512",
    year = "2015"
}

@article{Bezrukov:2009db,
    author = "Bezrukov, F. and Shaposhnikov, M.",
    title = "{Standard Model Higgs boson mass from inflation: Two loop analysis}",
    eprint = "0904.1537",
    archivePrefix = "arXiv",
    primaryClass = "hep-ph",
    doi = "10.1088/1126-6708/2009/07/089",
    journal = "JHEP",
    volume = "07",
    pages = "089",
    year = "2009"
}

@article{Bezrukov:2010jz,
    author = "Bezrukov, F. and Magnin, A. and Shaposhnikov, M. and Sibiryakov, S.",
    title = "{Higgs inflation: consistency and generalisations}",
    eprint = "1008.5157",
    archivePrefix = "arXiv",
    primaryClass = "hep-ph",
    doi = "10.1007/JHEP01(2011)016",
    journal = "JHEP",
    volume = "01",
    pages = "016",
    year = "2011"
}

@article{Bezrukov:2012sa,
    author = "Bezrukov, Fedor and Kalmykov, Mikhail Yu. and Kniehl, Bernd A. and Shaposhnikov, Mikhail",
    editor = "Moortgat-Pick, Gudrid",
    title = "{Higgs Boson Mass and New Physics}",
    eprint = "1205.2893",
    archivePrefix = "arXiv",
    primaryClass = "hep-ph",
    reportNumber = "DESY-12-074",
    doi = "10.1007/JHEP10(2012)140",
    journal = "JHEP",
    volume = "10",
    pages = "140",
    year = "2012"
}

@article{Enckell:2020lvn,
    author = {Enckell, Vera-Maria and Nurmi, Sami and R\"as\"anen, Syksy and Tomberg, Eemeli},
    title = "{Critical point Higgs inflation in the Palatini formulation}",
    eprint = "2012.03660",
    archivePrefix = "arXiv",
    primaryClass = "astro-ph.CO",
    reportNumber = "HIP-2020-33/TH",
    doi = "10.1007/JHEP04(2021)059",
    journal = "JHEP",
    volume = "04",
    pages = "059",
    year = "2021"
}

@article{Poisson:2023tja,
    author = "Poisson, Arthur and Timiryasov, Inar and Zell, Sebastian",
    title = "{Critical points in Palatini Higgs inflation with small non-minimal coupling}",
    eprint = "2306.03893",
    archivePrefix = "arXiv",
    primaryClass = "hep-ph",
    doi = "10.1007/JHEP03(2024)130",
    journal = "JHEP",
    volume = "03",
    pages = "130",
    year = "2024"
}

@article{Markkanen:2018bfx,
    author = "Markkanen, Tommi and Nurmi, Sami and Rajantie, Arttu and Stopyra, Stephen",
    title = "{The 1-loop effective potential for the Standard Model in curved spacetime}",
    eprint = "1804.02020",
    archivePrefix = "arXiv",
    primaryClass = "hep-ph",
    reportNumber = "IMPERIAL/TP/2018/TM/02, IMPERIAL-TP-2018-TM-02",
    doi = "10.1007/JHEP06(2018)040",
    journal = "JHEP",
    volume = "06",
    pages = "040",
    year = "2018"
}

@article{Branchina:2013jra,
    author = "Branchina, Vincenzo and Messina, Emanuele",
    title = "{Stability, Higgs Boson Mass and New Physics}",
    eprint = "1307.5193",
    archivePrefix = "arXiv",
    primaryClass = "hep-ph",
    doi = "10.1103/PhysRevLett.111.241801",
    journal = "Phys. Rev. Lett.",
    volume = "111",
    pages = "241801",
    year = "2013"
}

@article{Coleman:1973jx,
    author = "Coleman, Sidney R. and Weinberg, Erick J.",
    title = "{Radiative Corrections as the Origin of Spontaneous Symmetry Breaking}",
    doi = "10.1103/PhysRevD.7.1888",
    journal = "Phys. Rev. D",
    volume = "7",
    pages = "1888--1910",
    year = "1973"
}

@article{Gildener:1976ih,
    author = "Gildener, Eldad and Weinberg, Steven",
    title = "{Symmetry Breaking and Scalar Bosons}",
    reportNumber = "PRINT-76-0068 (HARVARD)",
    doi = "10.1103/PhysRevD.13.3333",
    journal = "Phys. Rev. D",
    volume = "13",
    pages = "3333",
    year = "1976"
}

@article{Neubert:2019mrz,
    author = "Neubert, Matthias",
    editor = "Davidson, Sacha and Gambino, Paolo and Laine, Mikko and Neubert, Matthias and Salomon, Christophe",
    title = "{Renormalization Theory and Effective Field Theories}",
    eprint = "1901.06573",
    archivePrefix = "arXiv",
    primaryClass = "hep-ph",
    reportNumber = "MITP/19-002",
    doi = "10.1093/oso/9780198855743.003.0001",
    month = "1",
    year = "2019"
}

@article{Callan:1970,
  title = {Broken Scale Invariance in Scalar Field Theory},
  author = {Callan, Curtis G.},
  journal = {Phys. Rev. D},
  volume = {2},
  issue = {8},
  pages = {1541--1547},
  numpages = {0},
  year = {1970},
  month = {Oct},
  publisher = {American Physical Society},
  doi = {10.1103/PhysRevD.2.1541},
  url = {https://link.aps.org/doi/10.1103/PhysRevD.2.1541}
}

@article{Symanzik:1970,
author="Symanzik, K.",
title="Small distance behaviour in field theory and power counting",
journal="Communications in Mathematical Physics",
year="1970",
month="Sep",
day="01",
volume="18",
number="3",
pages="227--246",
issn="1432-0916",
doi="10.1007/BF01649434",
url="https://doi.org/10.1007/BF01649434"
}

@article{Barvinsky:2009ii,
    author = "Barvinsky, A. O. and Kamenshchik, A. Yu. and Kiefer, C. and Starobinsky, A. A. and Steinwachs, C. F.",
    title = "{Higgs boson, renormalization group, and naturalness in cosmology}",
    eprint = "0910.1041",
    archivePrefix = "arXiv",
    primaryClass = "hep-ph",
    doi = "10.1140/epjc/s10052-012-2219-3",
    journal = "Eur. Phys. J. C",
    volume = "72",
    pages = "2219",
    year = "2012"
}

@article{Kadastik:2011aa,
    author = "Kadastik, Mario and Kannike, Kristjan and Racioppi, Antonio and Raidal, Martti",
    title = "{Implications of the 125 GeV Higgs boson for scalar dark matter and for the CMSSM phenomenology}",
    eprint = "1112.3647",
    archivePrefix = "arXiv",
    primaryClass = "hep-ph",
    doi = "10.1007/JHEP05(2012)061",
    journal = "JHEP",
    volume = "05",
    pages = "061",
    year = "2012"
}

@article{Okada:2015lia,
    author = "Okada, Nobuchika and Raut, Digesh",
    title = "{Running non-minimal inflation with stabilized inflaton potential}",
    eprint = "1509.04439",
    archivePrefix = "arXiv",
    primaryClass = "hep-ph",
    doi = "10.1140/epjc/s10052-017-4799-4",
    journal = "Eur. Phys. J. C",
    volume = "77",
    number = "4",
    pages = "247",
    year = "2017"
}

@article{Okada:2010jf,
    author = "Okada, Nobuchika and Rehman, Mansoor Ur and Shafi, Qaisar",
    title = "{Tensor to Scalar Ratio in Non-Minimal $\phi^4$ Inflation}",
    eprint = "1005.5161",
    archivePrefix = "arXiv",
    primaryClass = "hep-ph",
    doi = "10.1103/PhysRevD.82.043502",
    journal = "Phys. Rev. D",
    volume = "82",
    pages = "043502",
    year = "2010"
}

@article{Okada:2010jd,
    author = "Okada, Nobuchika and Shafi, Qaisar",
    title = "{WIMP Dark Matter Inflation with Observable Gravity Waves}",
    eprint = "1007.1672",
    archivePrefix = "arXiv",
    primaryClass = "hep-ph",
    doi = "10.1103/PhysRevD.84.043533",
    journal = "Phys. Rev. D",
    volume = "84",
    pages = "043533",
    year = "2011"
}

@article{Okada:2011en,
    author = "Okada, Nobuchika and Rehman, Mansoor Ur and Shafi, Qaisar",
    title = "{Non-Minimal B-L Inflation with Observable Gravity Waves}",
    eprint = "1102.4747",
    archivePrefix = "arXiv",
    primaryClass = "hep-ph",
    doi = "10.1016/j.physletb.2011.06.044",
    journal = "Phys. Lett. B",
    volume = "701",
    pages = "520--525",
    year = "2011"
}

@article{Burgess:2009ea,
    author = "Burgess, C. P. and Lee, Hyun Min and Trott, Michael",
    title = "{Power-counting and the Validity of the Classical Approximation During Inflation}",
    eprint = "0902.4465",
    archivePrefix = "arXiv",
    primaryClass = "hep-ph",
    reportNumber = "PI-PARTPHYS-121",
    doi = "10.1088/1126-6708/2009/09/103",
    journal = "JHEP",
    volume = "09",
    pages = "103",
    year = "2009"
}

\end{document}